\documentclass[sigconf, nonacm]{acmart}

\usepackage{balance}

\usepackage[ruled,linesnumbered]{algorithm2e} 

\usepackage{algorithmic}
\usepackage{float}
\usepackage{listings}
\begin{document}
%
\title{A Frequency-aware Software Cache for Large Recommendation System Embeddings}




\author{Jiarui Fang}
\affiliation{
\institution{HPC-AI Technology Inc.}            
}
\email{fangjr@hpcaitech.com}

\author{Geng Zhang}
\authornote{The work is done when Geng is at HPC-AI Technology Inc.}
\affiliation{
\institution{School of Computer Science, Wuhan University} 
}
\email{zhangg@whu.edu.cn}

\author{Jiatong Han}
\affiliation{
\institution{HPC-AI Technology Inc.}        
}
\email{jiatong.han@hpcaitech.com}

\author{ Shenggui Li, Zhengda Bian, Yongbin Li}
\affiliation{
\institution{HPC-AI Technology Inc.}        
}
\email{{lisg, bian.zhengda, ybl}@hpcaitech.com}

\author{Jin Liu}
\affiliation{
\institution{School of Computer Science, Wuhan University}            
}
\email{jinliu@whu.edu.cn}

\author{Yang You}
\authornote{corresponding author. Yang You is a faculty member at NUS; this work was done at HPC-AI Technology Inc.}
\affiliation{
\institution{National University of Singapore (NUS)}            
}
\email{youy@comp.nus.edu.sg}


\begin{abstract}
Deep learning recommendation models (DLRMs) have been widely applied in Internet companies. 
The embedding tables of DLRMs are too large to fit on GPU memory entirely.
We propose a GPU-based software cache approaches to dynamically manage the embedding table in the CPU and GPU memory space by leveraging the id's frequency statistics of the target dataset.
Our proposed software cache is efficient in training entire DLRMs on GPU in a synchronized update manner.
It is also scaled to multiple GPUs in combination with the widely used hybrid parallel training approaches.
Evaluating our prototype system shows that we can keep only 1.5\% of the embedding parameters in the GPU to obtain a decent end-to-end training speed.
\end{abstract}
\maketitle

%

\section{Introduction}
The DLRMs are important machine learning applications that offer a personalized user experience in industry~\cite{gomez2015netflix, naumov2019deep, smith2017two}.
DLRMs are typically based on both dense and sparse features.
The embedding table is used to handle sparse features.

Embedding tables are growing larger nowadays because an increase in embedding capacity usually leads to an improvement in recommendation quality. Currently, large embeddings can easily scale up to Trillions of parameters~\cite{lian2021persia}.
The embedding tables are too large to fit on GPU memory entirely.
Existing solutions resort to heterogeneous training typically involving CPU computation of the embedding part and GPU computation of the dense part.
However, the random embedding lookups on the CPU can be an order of magnitude slower than on that the GPU.
To hide the CPU computational overhead, They often introduce asynchronous update strategies, which are notorious for their uncertain convergence rate.

This work focuses on using homogeneous training of the entire model solely on GPU in a synchronous updating manner.
To meet huge memory requirements, we apply a software cache to dynamically move active embedding rows between CPU and GPU.
Different from the existing software cache approaches, like the Unified Virtual Memory (UVM) adopted in TorchRec, our cache leverages the dataset id frequency statistics to improve the cache hit ratio and applies a set of optimization to both reduce cache indexing overhead and improve PCI-e bandwidth utilization.
Our contributions are listed as follows.
\begin{itemize}
    \item We proposed a software cache approach to manage embedding parameters in hybrid memory space including CPU and GPU. Our cache-related operations are parallel executed on GPU. Therefore has a limited impact on the overall performance. The cache utilizes the id frequency statistic of the target dataset to improve the cache hit ratio.
    \item We scale the embedding using the cache to multiple GPUs using hybrid parallel training approaches.
    \item Our code is publicly available at url~\footnote{https://github.com/zxgx/FreqCacheEmbedding}.
\end{itemize}

\section{Related Work}
\subsection{Deep Learning Recommendation Model} 
Deep Learning Recommendation Model (DLRM)\cite{naumov2019deep} is a widely-adopted recommendation model that is able to take in both sparse and dense features that are essentially useful for predicting user-item matches. An embedding layer is used for sparse feature embedding look-ups, and our work centers around adapting embedding layer such that model training can be supported even with extremely large embedding sizes and relatively small GPU memory. There are also dense layers in DLRM that process dense inputs that are usually continuous, and an over-arch layer that processes combined results from dense layers and embedding layer. They will not be our focus in this work. 

\subsection{Large-Scale Embedding Training}
\subsubsection{Embedding Table Sharding}
There are multiple approaches to supporting giant embedding layer training in DLRM. One traditional and universally adopted approach is to use Tensor Parallelism (TP) that shards embedding tables in a specific manner (such as table-wise, or column-wise) onto multiple GPUs, and combines the look-up results using an all-reduce operation before proceeding with subsequent model steps. TorchRec\footnote{https://github.com/pytorch/torchrec}, as a recently released implementation of work~\cite{mudigere2022software}, has such functionality of embedding table sharding and TP.

\subsubsection{Hybrid CPU-GPU Embedding Storage}
Orthogonal to the above method, there are also works in distributing embedding tables onto GPU and CPU main memory, as GPU main memory is often expensive and restricted in size compared to CPU main memory. 
This implementation is often termed cached embedding or parameter server embedding by recent papers. 
Persia \cite{lian2021persia} designs a parameter server that resides in CPU memory and communicates and manages the embedding weights updates informed by model training happening in GPU. 
It is shown to be able to support trillions of parameters training. There have also been works \cite{ginart2021mixed, adnan2021accelerating, miao2021het} that observe an empirical long-tail distribution of look-up queries, designing frequency-aware cache strategies that share a similar underlying motivation to ours. 
Hotline \cite{adnan2022heterogeneous} makes effort to accelerate CPU-GPU communication by designing a specialized hardware scheduler that prioritizes more frequently queried embeddings and delays less frequent ones.

Specific to DLRM, there are both embedding layers and dense layers, which are I/O-intensive and compute-intensive respectively. Persia \cite{lian2021persia} put all embedding querying duty on CPU while putting dense layers computation duty entirely on GPU, which in this paper is termed as heterogeneous training. Compared to homogeneous training that uses GPU or CPU only, heterogeneous training can better leverage the CPU and GPU resources. We further this effort and put frequently queried embedding cache in GPU and use GPU parallelized operations to speed up embedding look-ups. 

\subsubsection{Synchronous Embedding Updates}
During actual model training, embedding weights are updated with each training epoch ending synchronously. One simple idea to speed up training is to update embedding weights asynchronously and overlap the step with the next training step as implemented in HET \cite{miao2021het}. However, we take a synchronous approach finally to avoid uncertainty convergence and possible model accuracy loss.

\section{Motivations}
Heterogeneous training is the most popular way to meet excessive memory requirements for the training of embeddings.
Because embedding tables are data-intensive operations that featured large memory requirements and small computation requirements,
previous work~\cite{miao2021het, lian2021persia, adnan2022heterogeneous} proposed to execute the embedding tables on CPU, and execute the remainder dense DNN part on GPU.
In this way, communication between CPU and GPU via PCI-e is the main performance bottleneck.
Work~\cite{miao2021het, lian2021persia} heavily relies on asynchronous update strategies to 
overlap the computation and data transfer.
The asynchronous updates introduce uncertainty to the convergence rate of the training, which is unfriendly to the system users.

\begin{figure}[ht!]
\centering
\includegraphics[width=0.5\textwidth]{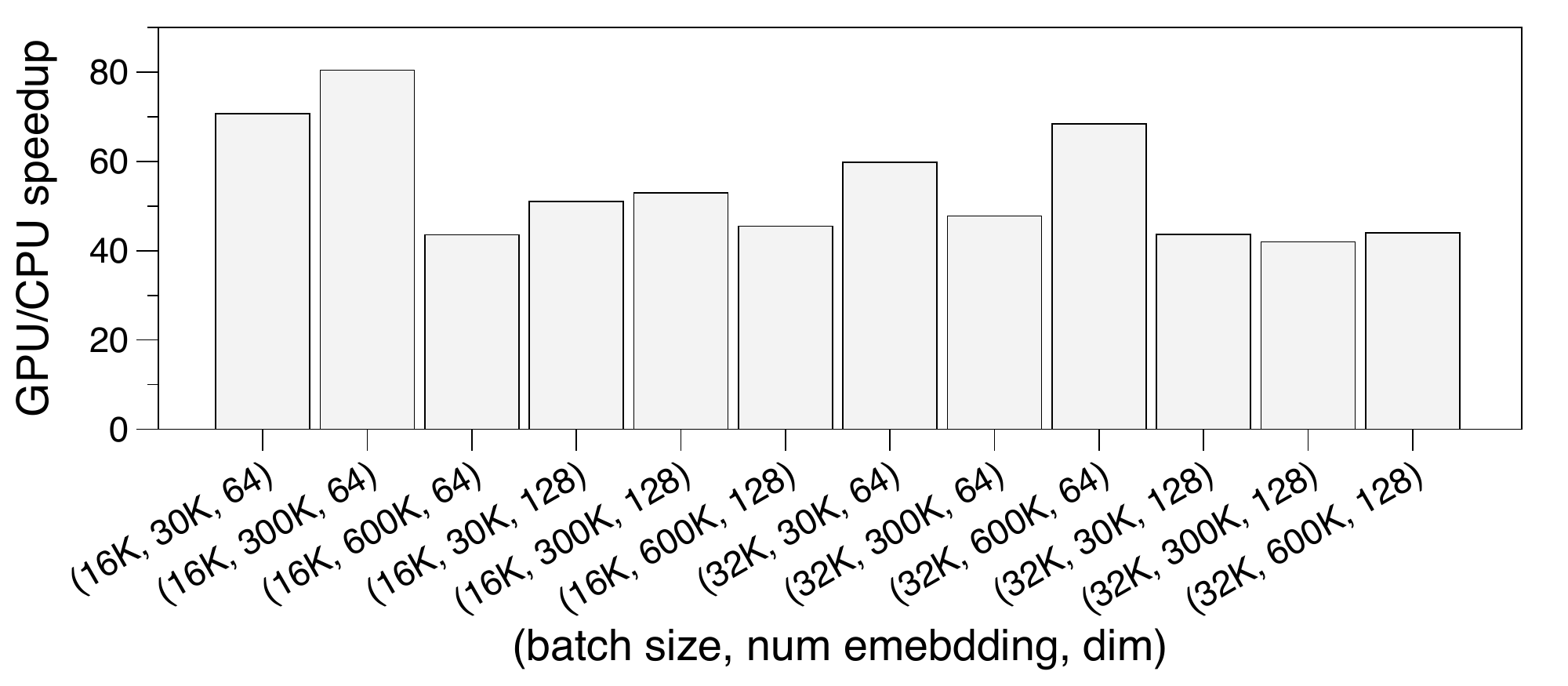}
\caption{Speedup of GPU to CPU on EmbeddingBags execution using PyTorch.}
\label{fig:embeddingbench}
\end{figure}

This work focuses on applying the homogeneous training approaches to training embedding tables entirely on GPU.
We observe that running embedding on GPU has significant gains compared to CPU.
Due to the irregular memory access pattern of embedding lookup, 
the hardware cache on the CPU benefits little from the training embedding tables.
Therefore, the memory access speed of embedding is largely limited by the bandwidth of the main memory bus,
and the bandwidth of High Bandwidth Memory (HBM) on GPU is orders of magnitude higher than the bandwidth of DDR4 or DDR5 memory on CPU.
Figure~\ref{fig:embeddingbench} shows around 50x speedup on GPU to CPU executing PyTorch \texttt{EmbeddingBag}.
In the benchmark, the CPU is AMD EPYC 7543 32-Core processor and the GPU is NVIDIA A100.
From the above benchmark data, it is implied that even if there are 10\% cache misses during the training process, there will be a 500\% increase in computing latency even if we still do not consider the data transmission.

\begin{figure*}[ht]
\centering
\includegraphics[width=0.85\textwidth]{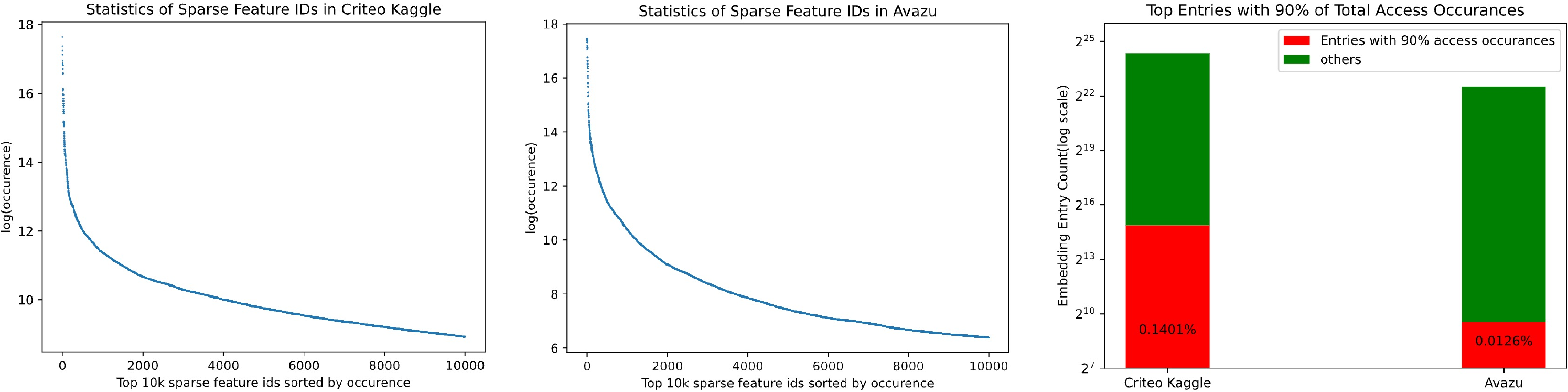}
\caption{The id frequency distribution in Criteo Kaggle and Avazu dataset. The left two figures show the frequency distribution for top-10K ids. 
The number in the right figure show the percentage of ids accounting for the top-10\% access.}
\label{fig:freqstars}
\end{figure*}

Software cache is necessary considering the large embedding table sizes (usually tens of GB, at most tens of TB) that can hardly fit into GPU memory. With such an approach, CPU memory serves as the main embedding storage that holds all of the embedding data, while the GPU only stores a small faction of the embedding data to be involved in computation in the near future. With similar motivation, TorchRec adopts a Unified Virtual Memory (UVM)~\footnote{https://developer.nvidia.com/blog/unified-memory-cuda-beginners/} as the software cache. It provides a single, unified memory address space for CPU and GPU and automatically replaces and evicts unused pages. However, this approach brings extra overhead from the data transmission process which operates at the granularity of embedding rows, leaving the bandwidth of PCI-e not sufficiently utilized. With the extra overhead, the software cache approach UVM is not ideal in its cost efficacy, and thus insufficiently good for DLRM. And as a result, some work resorts to hardware solutions to implement a fast caching approach~\cite{mudigere2022software, adnan2022heterogeneous}.

\textbf{The question is: Can we implement an efficient software cache on embeddings to implement homogeneous GPU compute for DLRMs without extra hardware accelerators?}
We solve the performance issue of the software cache from two viewpoints. 
First, increase the cache hit ratio.
It has been reported that there exist skewed popularity distributions of embeddings~\cite{miao2021het}.
As shown in the Figure~\ref{fig:freqstars}, a very small number of popular ids occur very often.
For example, the top 0.14\% and top 0.012\% popular embedding rows in the Criteo dataset and Avazu account for 90\% total number of updates.
The observation motivates us to reduce the communication by caching frequently updated embeddings rows into the GPU memory.
Second, reduce caching overhead, including cache key indexing and data transmission cost. 
To reduce key indexing overhead, we design a parallel caching method that gets executed on multiple GPUs. 
To increase the bandwidth utilization of data transmission, we increase the message size by converting transmission from row-wise to block-wise into the cache.

\section{The Software Cache}
\begin{figure}[ht!]
\centering
\includegraphics[width=0.5\textwidth]{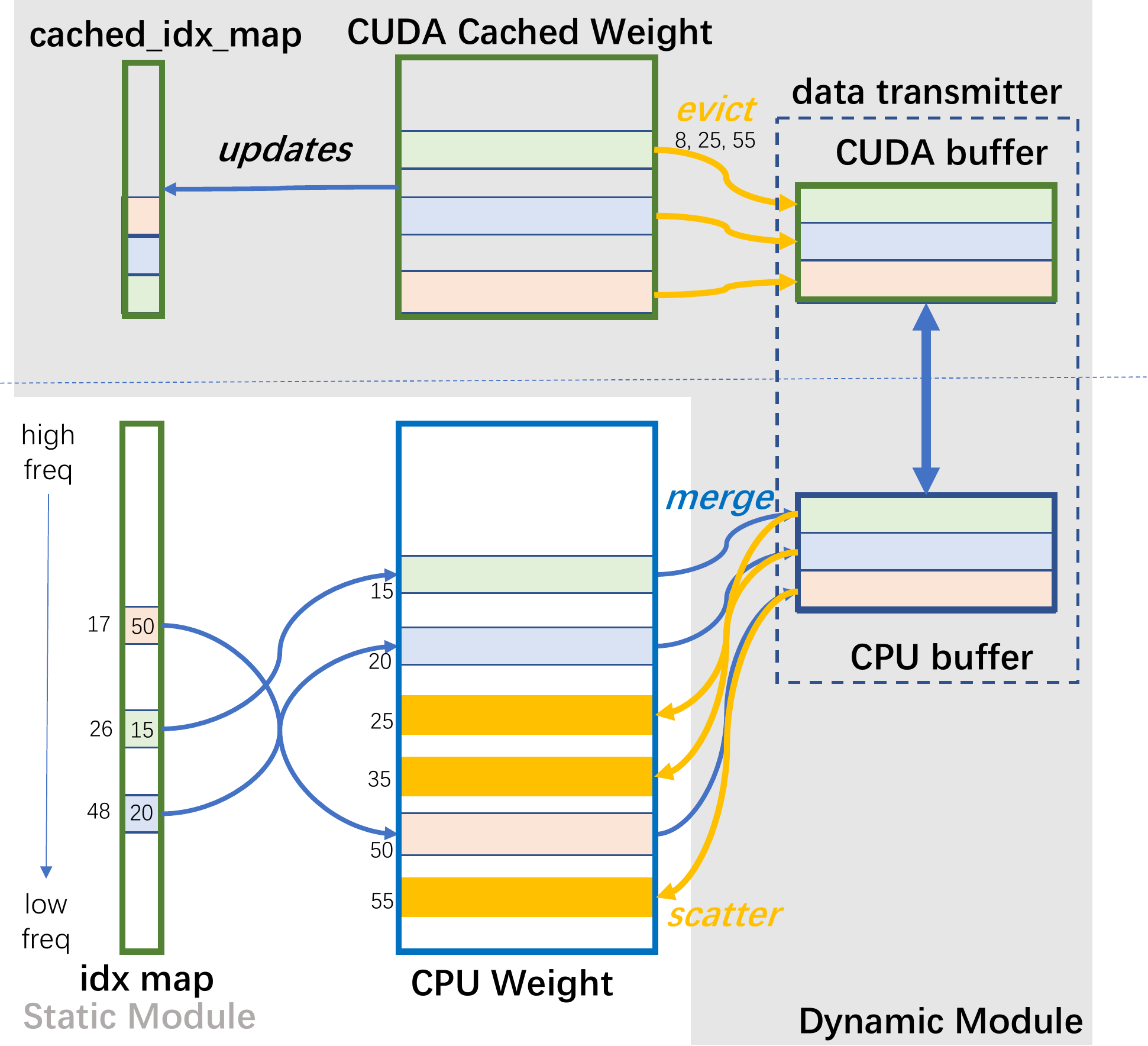}
\caption{The Framework of the Frequency-aware Software Cache.}
\label{fig:framework}
\end{figure}

We design a frequency-aware software cache on GPU leveraging id frequency in the dataset as shown in Figure~\ref{fig:framework}. 
It is a two-level cache with uses the hybrid memory space composed of CPU and GPU memory.
The embedding tables are resident in CPU memory as \textit{CPU weight}, while some copies of embedding rows are placed in GPU memory as CUDA \textit{Cached Weight}.
During DLRMs training, the embedding rows activated by the input ids of this iteration are dynamically transmitted from CPU Weight to CUDA Cached Weight.
If necessary, it uses the Least Frequently Used (LFU) algorithm to evict unused rows.

The software cache is innovative in the following aspects.
First, it uses the frequency statistics of the dataset id to improve the hit rate.
These statistics are prepared into the cache statically before training.
Secondly, it uses the buffer mechanism to improve the bandwidth utilization of data in PCI-e transmission, and also strictly limits the memory used.
Finally, cache-related operations are parallelized on GPU and introduces very little overhead.

\subsection{Terminology}
Before introducing the cache algorithm, let's define some terms.
\begin{itemize}
    \item \textbf{ids}: The ids from dataset.
    \item \textbf{cpu\_row\_idx}: The index of the embedding row in the CPU Weight.
    \item \textbf{gpu\_row\_idx}: The index of the embedding row corresponding in CUDA Cached Weight.
    \item \textbf{idx\_map}: a map from id to cpu\_row\_idx.
    \item \textbf{cache\_idx\_map}: a map from gpu\_row\_idx to cpu\_row\_idx.
\end{itemize}

\subsection{Static Module}
The static module of the system is prebuilt before training that does not change afterward. 
The static module contains the CPU Weight, which contains the entire original embedding weight values but is reordered in rows from high to low according to its occurrence frequency in the target dataset.
In the original weight, the input id is the same as the row number of the weight. 
This does not hold for reorder CPU weight. 
Therefore, it uses idx\_map, implemented with a 1D array, to convert the input id to the reordered row number. 

To get the id frequency distribution, we can simply scan the dataset before training.
If the dataset is too large, we can use the sampling method proposed in work~\cite{adnan2021accelerating}.

\subsection{Dynamic Module}
The dynamic module of the system keeps updating during the training according to input ids. 
It transmits embedding rows between CPU and GPU efficiently to place the embedding data required by each iteration in the CUDA memory before computation.
The dynamic module consists of the CUDA cached weight, a cache\_idx\_map, and a data transmitter. 

The CUDA Cached Weight contains a fraction of the CPU weight.
The cached\_idx\_map, also a 1D array, converts the cpu\_row\_idx of the CPU Weight to the gpu\_row\_idx of the CUDA Cached Weight.
The ratio of the size CUDA Cached Weight to the size of CPU Weight is named as \textbf{cache ratio}, which can be 1.5\% by default.

The data transmitter is responsible for the bidirectional transmission of data between the CUDA Cached Weight and the CPU Weight. 
It uses a buffering approach to solve the problem of insufficient utilization of PCI-e bandwidth for fine-grained row-wise transmission.
The embedding rows scattered in the memory are concentrated as continuous data blocks in the local memory of the source device.
Then the blocks are transferred between the CPU and GPU and scattered to the corresponding position of the target memory. 
Moving data in blocks can improve the PCI-e bandwidth utilization. 
The overhead of concentration and scattering is also very small.
Because the data transfer bandwidth on the CPU and GPU is several orders of magnitude higher than that of PCI-e.

The memory footprint of the data transmitter is strictly limited.
In the worst case, all input ids are cache missed. 
We need to transfer $batch\_size \times feature\_num \times embedding\_size$ elements.
In order to prevent the buffer from occupying too much memory, we strictly limit the buffer size. 
If the transferred data is larger than the buffer, we complete the transfer multiple times.

All cache-related operations are executed on the GPU.
In Figure~\ref{fig:framework}, all the green boxes are stored in CUDA memory and we use the GPU parallel operations to implement cache-related operations.
The workflow of the dynamic module during the training process is shown in the Algorithm~\ref{alg:preparecache}. 
First, we identify the cpu\_row\_idx in the CPU Weight activated by the current batch of input ids of this iteration. 
Then, we find out the CPU embedding rows that are not in CUDA Cached Weight.
We need to move these rows from the CPU to the GPU. 
To ensure that these rows have enough space to be placed in the CUDA Cached Weight, 
we may need to evict some rows from the CPU Cached Weight back to the CPU Weight through the data transmitter. 
Then, the data transmitter moves the embedding rows to the CUDA Cached Weight and updates the corresponding entries in the cached\_idx.
The cache indexing operations, involving \textbf{unique}, \textbf{isin}, \textbf{nonzero}, \textbf{index\_copy}, \textbf{index\_fill} and \textbf{argsort}, are executed on GPU in parallel using highly optimized APIs provided by PyTorch.
After cache-related operations, Deep Learning frameworks, PyTorch in our case, uses the CUDA Cached Weight to finish the training process of this iteration on GPU.
 
\begin{algorithm}
\small
\caption{Cache-related Operations}
\label{alg:preparecache}
\SetKwProg{func}{Function}{}{end}
\func{PrepareCache(ids)}{
    \state /*Convert ids from dataset to gpu\_row\_idxs in Cached CUDA Weight*/\\
    \state \textbf{Require:} idx\_map, cached\_idx\_map;\\
    \state cpu\_row\_idxs = \textbf{unique}(\textbf{index\_select}(idx\_map, dim=0, index=ids));\\
    \state comm\_cpu\_row\_idxs = idx\_map[\textbf{isin}(cpu\_row\_idxs, cached\_idx\_map, invert=True)];\\
    \state /*transmitting embeddings row between CPU and GPU.*/\\
    \state \textbf{PrepareCPURowIdxOnGPU}(comm\_cpu\_row\_idxs, cpu\_row\_idxs);\\
    \state gpu\_row\_idxs= \textbf{index\_select}(cached\_idx, dim=0, cpu\_row\_idxs);\\
    \state \textbf{Return} gpu\_row\_idxs;\\
}
\func{PrepareCPURowIdxOnGPU(cpu\_row\_idxs, backlist)}{
    \state \textbf{Require:} cache\_capacity;\\
    \state \textbf{Require:} cache\_idx;\\
    \state \textbf{Require:} data\_transmitter;\\
    \state evict\_num = \textbf{numel}(cpu\_row\_idxs) - cache\_capacity;\\
    \If{evict\_num $>$ 0:}{
        \state /*make sure cpu\_row\_idx in backlist stay in GPU and won't be evicted*/\\
        \state mask\_cpu\_row\_idx = \textbf{isin}(cache\_idx\_map, backlist);\\
        \state /*set ids w.r.t the mask\_idx in cache\_idx to -2 (less than any other ids)*/\\
        \state backup\_idxs = cache\_idx\_map[mask\_cpu\_row\_idx].clone();\\
        \state invalid\_idx = \textbf{nonzero}(backup\_gpu\_row\_idx).squeeze(1);\\
        \state cache\_idx\_map.\textbf{index\_fill\_}(dim = 0, index=invalid\_idx, value = -2);\\
        \state /*select the max evict\_num ids to be evicted from GPU*/\\
        \state evict\_gpu\_row\_idxs = \textbf{argsort}(cache\_idx\_map, descending=True)[:evict\_num];\\
        \state /*revert value of cache\_idx\_map back to their origin value*/\\
        \state cached\_idx\_map.\textbf{index\_copy\_}(0, index=invalid\_idx, value = backup\_idxs);\\
        \state target\_cpu\_row\_idx = cached\_idx\_map[evict\_idx];\\
        \state /*move rows of evict\_idx from Cached CUDA Weight to CPU weight*/\\
        \state \textbf{data\_transmitter}.move\_to\_cpu(evict\_gpu\_row\_idxs, target\_cpu\_row\_idx);\\
    }
    \state /*-1 in cache\_idx indicates id not in GPU*/\\
    \state target\_gpu\_row\_idxs = \textbf{nonzero}(cache\_idx\_map == -1).squeeze(1)[:\textbf{numel}(ids)]
    \textbf{data\_transmitter}.move\_to\_gpu(cpu\_row\_idx, target\_gpu\_row\_idxs);\\
    \state cache\_idx[target\_gpu\_row\_idxs] = idcpu\_row\_idxs;\\
}
\end{algorithm}

We design an LFU eviction strategy to evict the current not-in-use rows to the CPU weight. 
Different from the traditional LFU method, we use the id frequency statistics of the dataset collected in advance. 
Suppose we need to evict $n$ rows. 
We only need to select the $n$ largest cpu\_row\_idx, and then evict out the corresponding rows.
The evicted rows are also transmitted via the data transmitter.
Cache warm-up can be placed before training.
We fill in the CUDA Cached Weight using the highest frequency ids.

\subsection{Scaling to multiple GPUs}
\begin{figure}[ht!]
\centering
\includegraphics[width=0.45\textwidth]{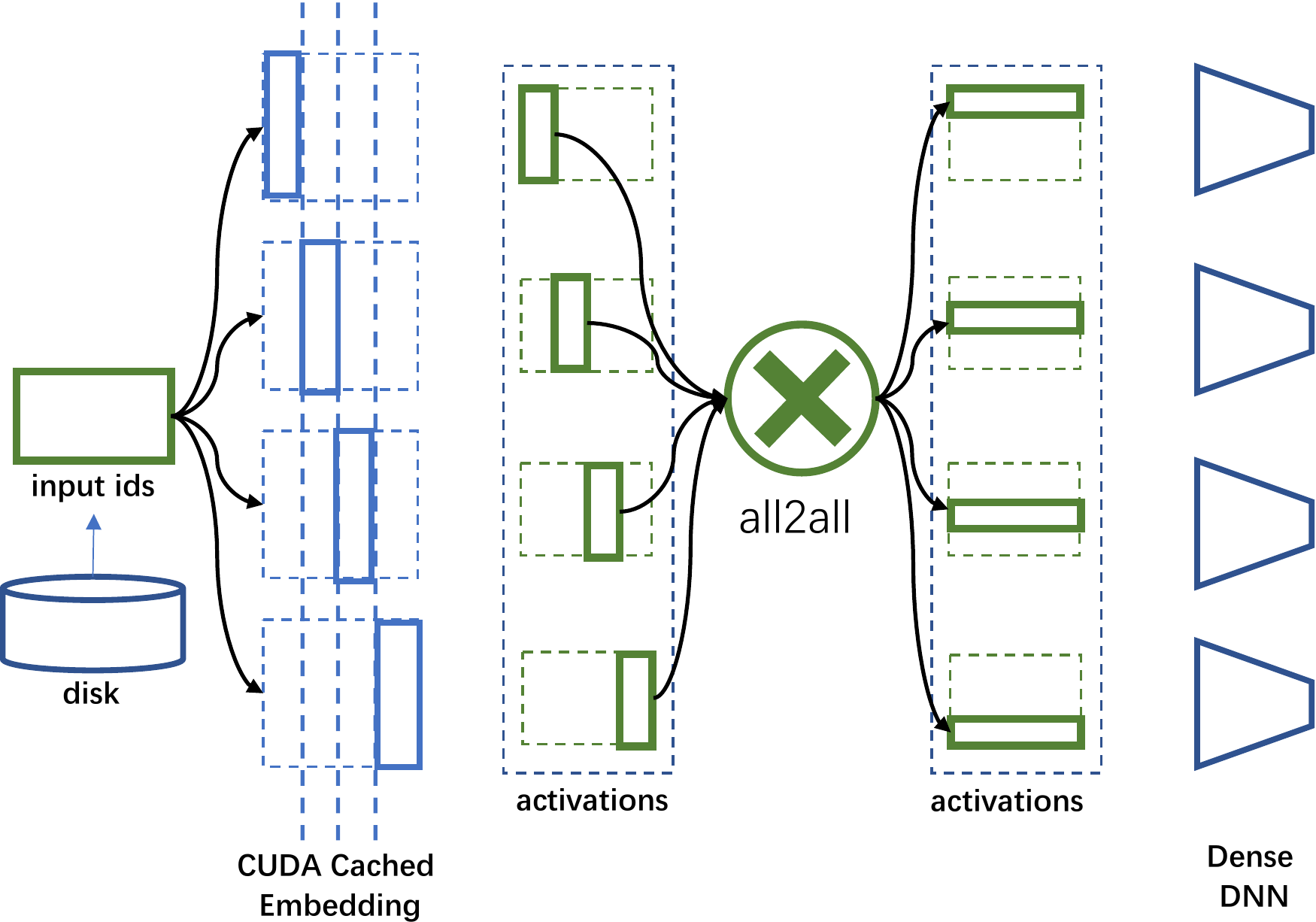}
\caption{Hybrid Parallel Approach to Scale Freq-aware Embedding on 4 GPUs.}
\label{fig:parallel}
\end{figure}

Our embedding tables implemented with our frequency-aware software cache are compatible with the parallel training approaches~\cite{bian2021colossal}.
In this work, we apply the widely used hybrid parallel method to scale to multiple GPUs as shown in Figure~\ref{fig:parallel}. 
Tensor Parallel is used for the Embedding layer, and data parallel is used for the dense layers.
Switching the parallel mode requires additional communication operations. 
We need to perform all2all operations on the output activations.

\section{Experiments}
\subsection{Dataset \& Software \& TestBed}
Table~\ref{tab:expr} reports the hardware of our testbed.
The GPUs are connected via a PCI-e 3.0 switch.
For Criteo Kaggle dataset, we directly use the 13 continuous features as the dense features for each sample and construct the embedding table for each of the 26 sparse features.
Due to the seriously imbalanced labels, we found that it requires non-trivial efforts to achieve convergence for DLRM trained on Avazu dataset. 
To solve this issue, we adopt the preprocessing steps \footnote{https://www.kaggle.com/code/leejunseok97/deepfm-deepctr-torch} commonly suggested by the Kaggle community.
It is worth mentioning that we focus on the effectiveness of the cache-based training scheme. 
Therefore, we focus on the consistency of the convergence of the baseline and our improved system, rather than obtaining an optimal model accuracy.
The training/validation/test sets take 90\%/5\%/5\% of the whole dataset respectively.
The statistics of the preprocessed datasets are summarized in Table~\ref{tab:dataset}.

\begin{table*}[ht!]
\small
\caption{Statistics of datasets}
\begin{center}
    \begin{tabular}{c|cccccc}
    \hline
         \textbf{Dataset}   & \textbf{\#Sparse feature}  & \textbf{\#Dense feature}   & \textbf{\#Embedding item}  & \textbf{\#Train}   & \textbf{\#Val} & \textbf{\#Test}    \\
    \hline
         Criteo Kaggle      &       26                  &       13                  &       33,762,577          &   39,291,954      &   3,274,331   &   3,274,332       \\
    \hline
         Avazu              &       13                  &       8                   &       9,445,823           &   36,386,071      &   2,021,448   &   2,021,448       \\
    \hline
    \end{tabular}
\end{center}
\label{tab:dataset}
\end{table*}

The embedding dimensions for all the embedding tables of DLRM are set as 128. 
The intermediate hidden sizes of bottom MLP for dense features are set to 512,256,128, and the top MLPs are of hidden sizes 1024,1024,512,256,1.
The learning rates are set as constants, 1 and 5e-2 for Criteo Kaggle and Avazu respectively.

We prototype our system using PyTorch v1.11.
Datasets include Criteo Kaggle and Avazu.
Software-wise we use TorchRec DLRM as the baseline.
The baseline uses a planner (an instance of EmbeddingShardingPlanner in TorchRec~\footnote{https://github.com/pytorch/torchrec/blob/main/examples/sharding/uvm.ipynb}) to automatically search the parallel strategies for embedding tables based on the hardware network topology.
According to our observation, the planner is able to search Embedding tables applied with UVM.
In our test cases, the table-wise parallel strategy is always the parallel strategy of the planners.
It means that each process holds different embedding tables, and may result in a potential memory load imbalance.
In contrast, our prototype system concatenates all embedding tables into a large embedding table and uses tensor parallelism to evenly partition the matrix along the embedding dimension, a.k.a column-wise 1D Tensor Parallel.
More parallel strategies for embedding will be investigated in our future work.

\begin{table}[ht!]
\small
\caption{Hardware Configuration for Experiments}
\begin{center}
\begin{tabular}{c|cccc}
\hline
\textbf{Device} &  \textbf{Architect} &  \textbf{Memory} & \\
\hline
8xGPU &  NVIDIA A100 & 80GB &  \\
\hline
CPU &  AMD EPYC 7543 32-Core  &  512GB \\
\hline
\end{tabular}
\end{center}
\label{tab:expr}
\end{table}

\subsection{Performance on one GPU}

\subsubsection{Accuracy}
\begin{figure}[ht!]
\centering
\includegraphics[width=0.45\textwidth]{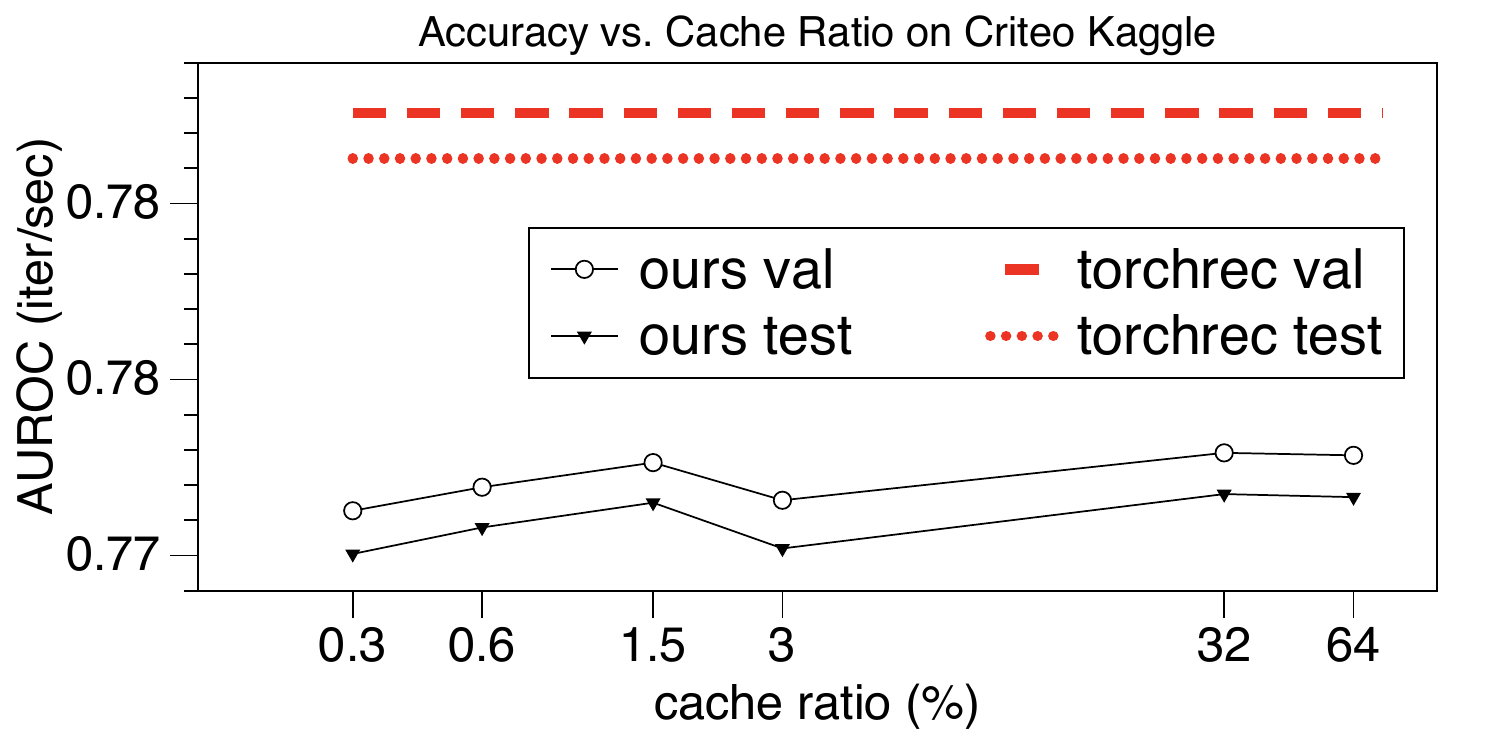}
\caption{Accuracy of DLRM on Criteo Kaggle vs. Cache Ratio using batch size as 16K on 1 GPU.}
\label{fig:criteo_acc}
\end{figure}

\begin{figure}[ht!]
\centering
\includegraphics[width=0.45\textwidth]{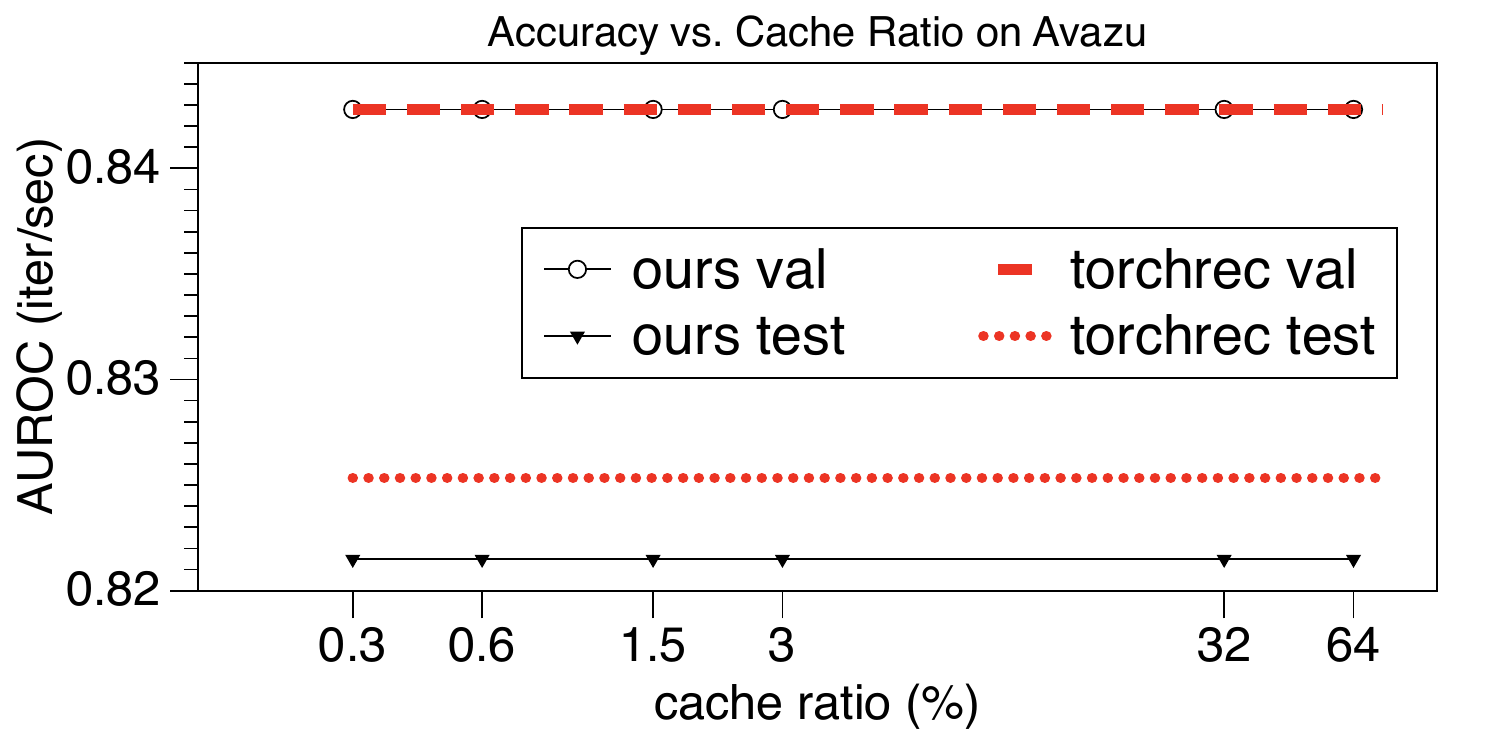}
\caption{Accuracy of DLRM on Avazu vs. Cache Ratio using batch size as 64K on 1 GPU.}
\label{fig:avazu_acc}
\end{figure}

As shown in Figure~\ref{fig:criteo_acc} and Figure~\ref{fig:avazu_acc}, we compare the AUROC achieved after 1 epoch training on validation dataset and test dataset, respectively.
The software cache using various cache ratio settings has minimal effect on the accuracy of the model.
The differences between our system and the baseline in AUROC metrics are less than 0.01 on the two datasets.

\subsubsection{Memory Usage}
\begin{figure}[ht!]
\centering
\includegraphics[width=0.45\textwidth]{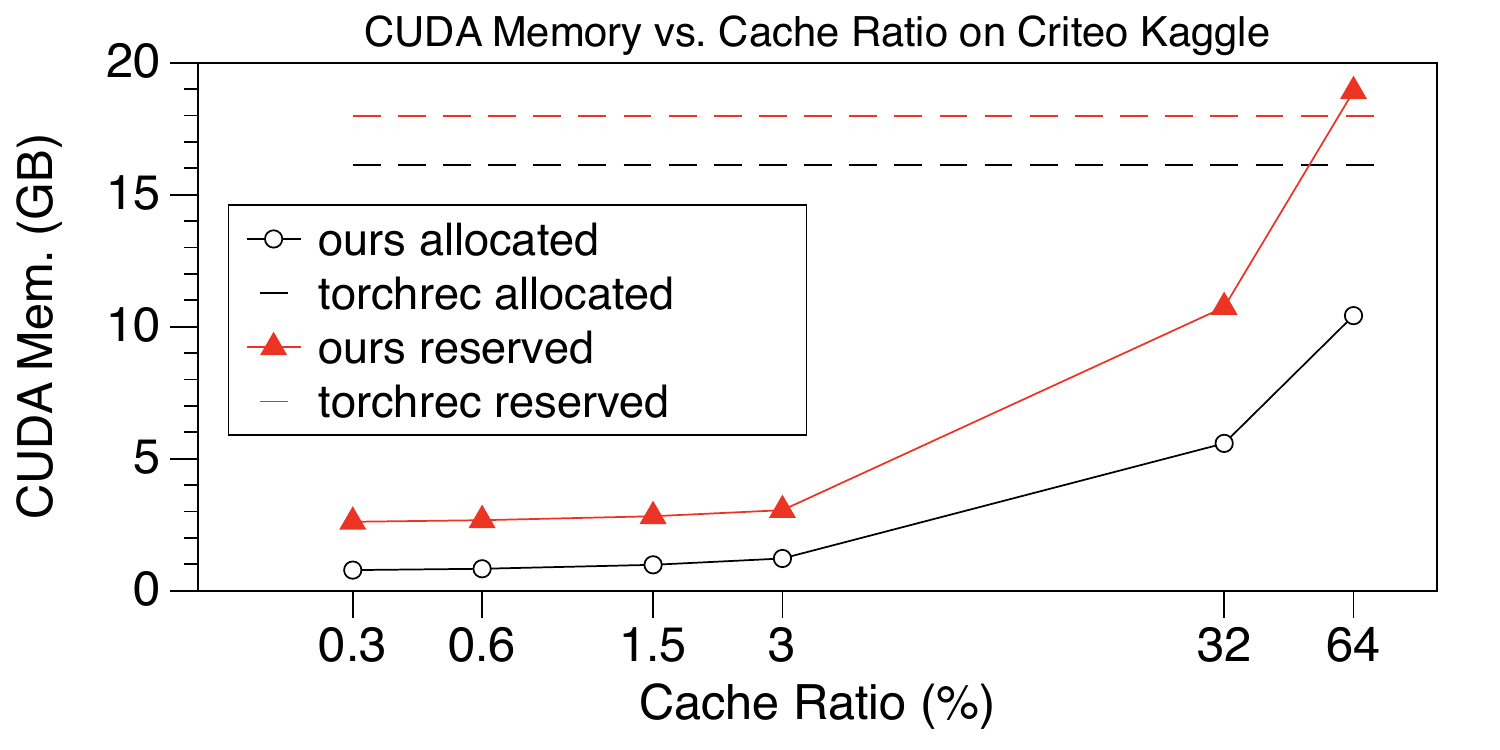}
\caption{Memory Usage of DLRM on Criteo Kaggle using batch size as 16K on 1 GPU.}
\label{fig:criteo_mem}
\end{figure}

\begin{figure}[ht!]
\centering
\includegraphics[width=0.45\textwidth]{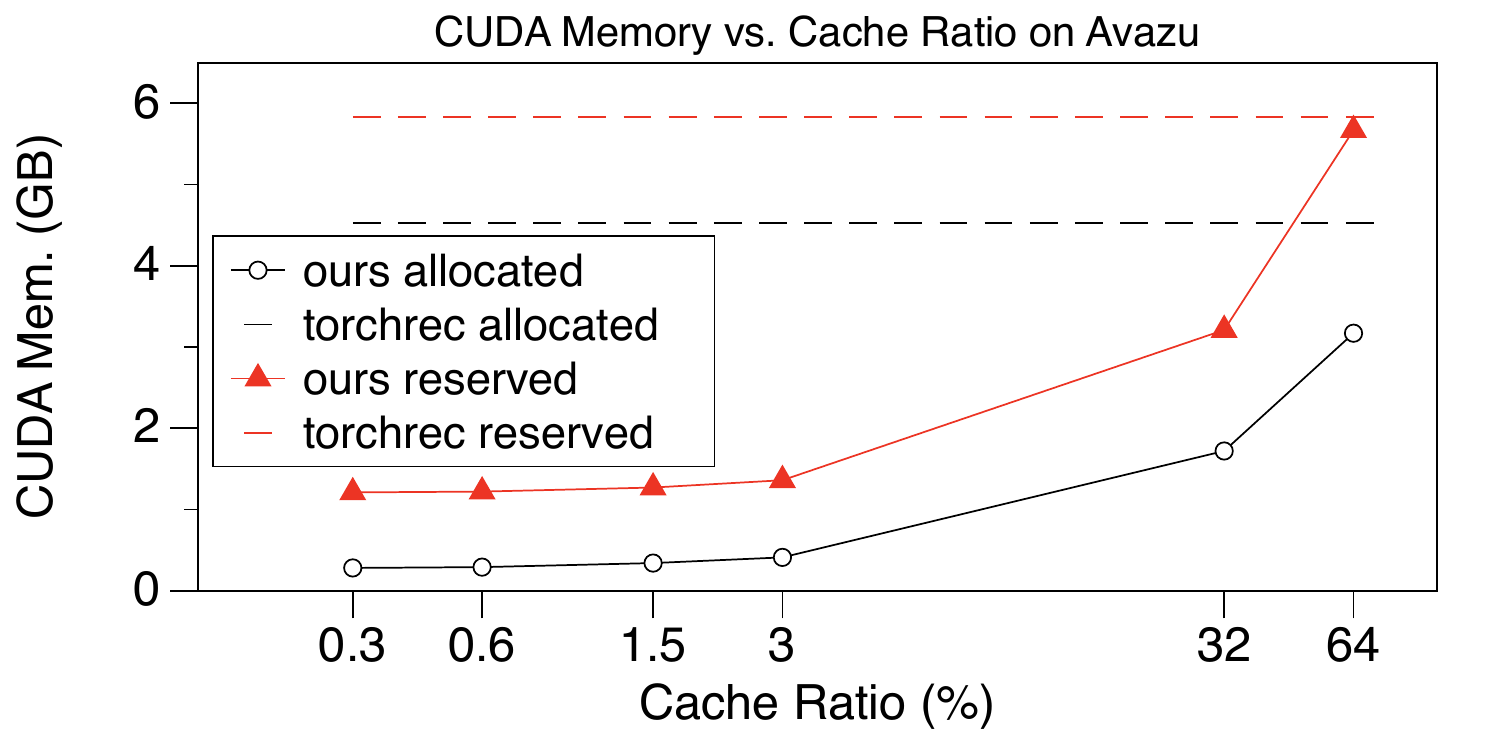}
\caption{Memory Usage of DLRM on Avazu using batch size as 64K on 1 GPU.}
\label{fig:avazu_mem}
\end{figure}

Figure~\ref{fig:criteo_mem} and Figure~\ref{fig:avazu_mem} present the CUDA memory usage of our system and TorchRec.
When setting the cache ratio to 1.5\%, our software cache can save around 80\% CUDA memory consumption.

\subsubsection{Throughput}
\begin{figure}[ht!]
\centering
\includegraphics[width=0.45\textwidth]{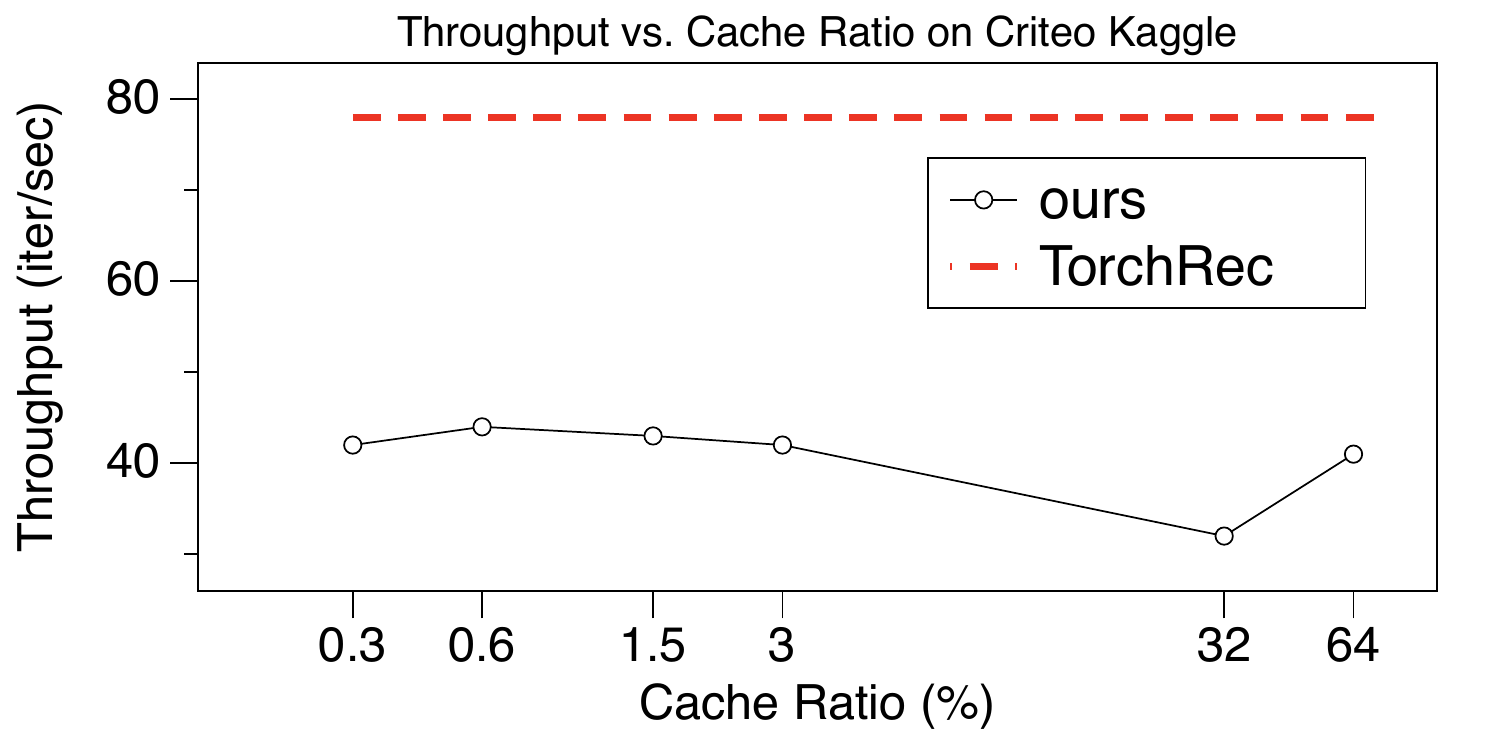}
\caption{Throughput of DLRM on Criteo Kaggle using global batch size as 16K on 1 GPU.}
\label{fig:criteo_cache_ratio}
\end{figure}

\begin{figure}[ht!]
\centering
\includegraphics[width=0.45\textwidth]{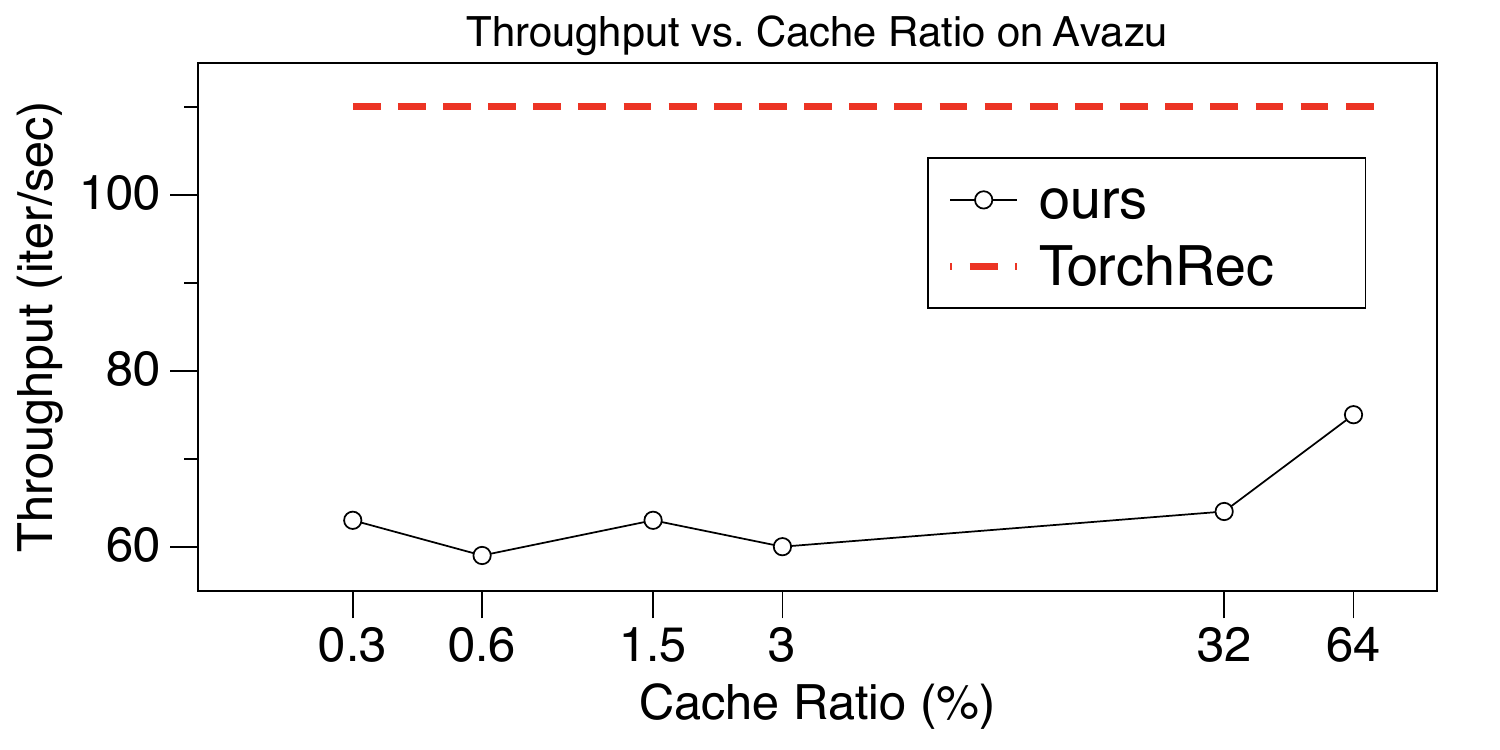}
\caption{Throughput of DLRM on Avazu vs. Cache Ratio using global batch size as 64K on 1 GPU.}
\label{fig:avazu_cache_ratio}
\end{figure}

Figure~\ref{fig:criteo_cache_ratio} and Figure~\ref{fig:avazu_cache_ratio} present how the cache ratio affects the DLRM performance on 1 GPU.

As can be seen from the figure, when the cache is 1.5\%, a nearly-optimal throughput can be obtained.
This is because when the cache ratio is higher than 1.5\%, more unused records are being loaded onto GPU, thus it does not help much to improve the hit rate. Instead, using a small cached CUDA weight can speed up the cache-related operations on GPU.

In the context of synchronous parameter updating, using software cache decrease by about 50\% of the overall performance compared to TorchRec.
It is implied that in order to make training with larger embeddings possible, some performance degradation was acceptable.

\subsection{Performance on multiple GPUs}

\subsubsection{Memory Usage}
\begin{figure}[ht!]
\centering
\includegraphics[width=0.45\textwidth]{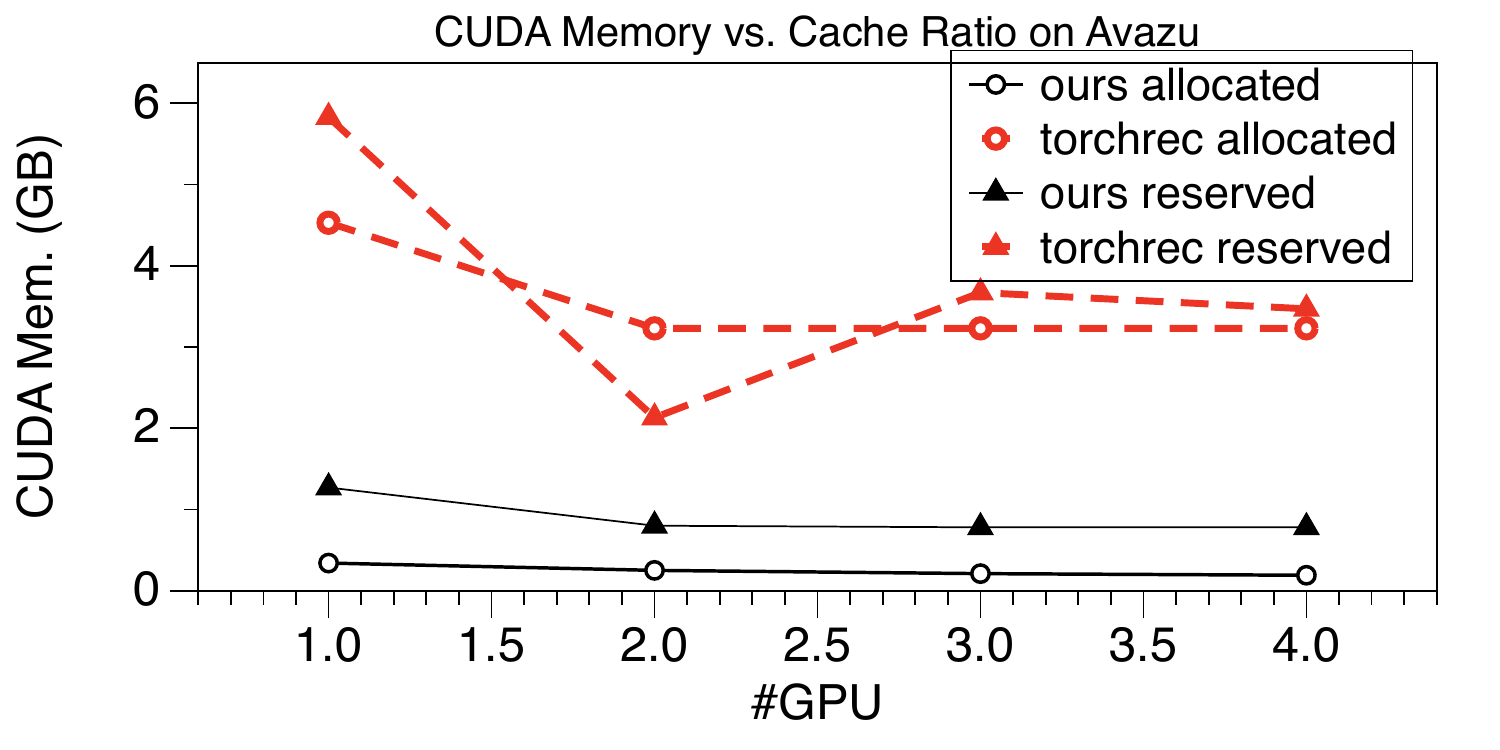}
\caption{Memory Usage of DLRM on Avazu when scaling to 8 GPUs.}
\label{fig:avazu_mem_mgpu}
\end{figure}

\begin{figure}[ht!]
\centering
\includegraphics[width=0.45\textwidth]{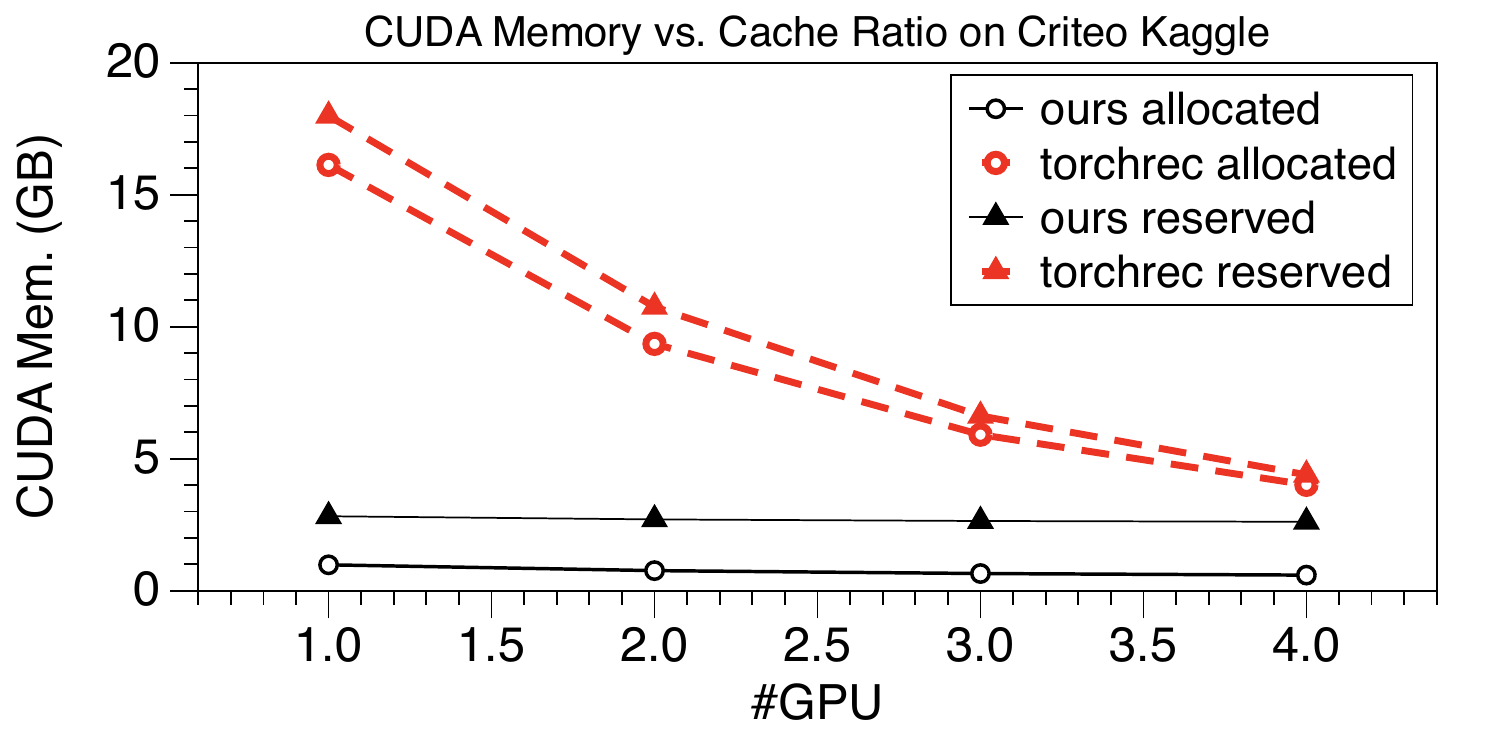}
\caption{Memory Usage of DLRM on Criteo Kaggle when scaling to 8 GPUs.}
\label{fig:criteo_mem_mgpu}
\end{figure}

Figure~\ref{fig:criteo_mem_mgpu} and Figure~\ref{fig:avazu_mem_mgpu} present the memory usage when we scale model training from 1 GPU to 8 GPUs.
Our GPU memory usage is still significantly less than TorchRec on multiple GPUs.
Also, the splitting of the Embeddingbag by TorchRec leads to unbalanced GPU memory usage. 
We make the maximum memory usage when experimenting with TorchRec on a single GPU.
By increasing the degree of parallelism, TorchRec can reduce memory usage per GPU, but still requires more than ours.
Since our software caching method has already used very little memory on a single GPU, no obvious memory savings are observed and expected when extended to multiple GPUs.

\subsubsection{Throughput}
\begin{figure}[ht!]
\centering
\includegraphics[width=0.45\textwidth]{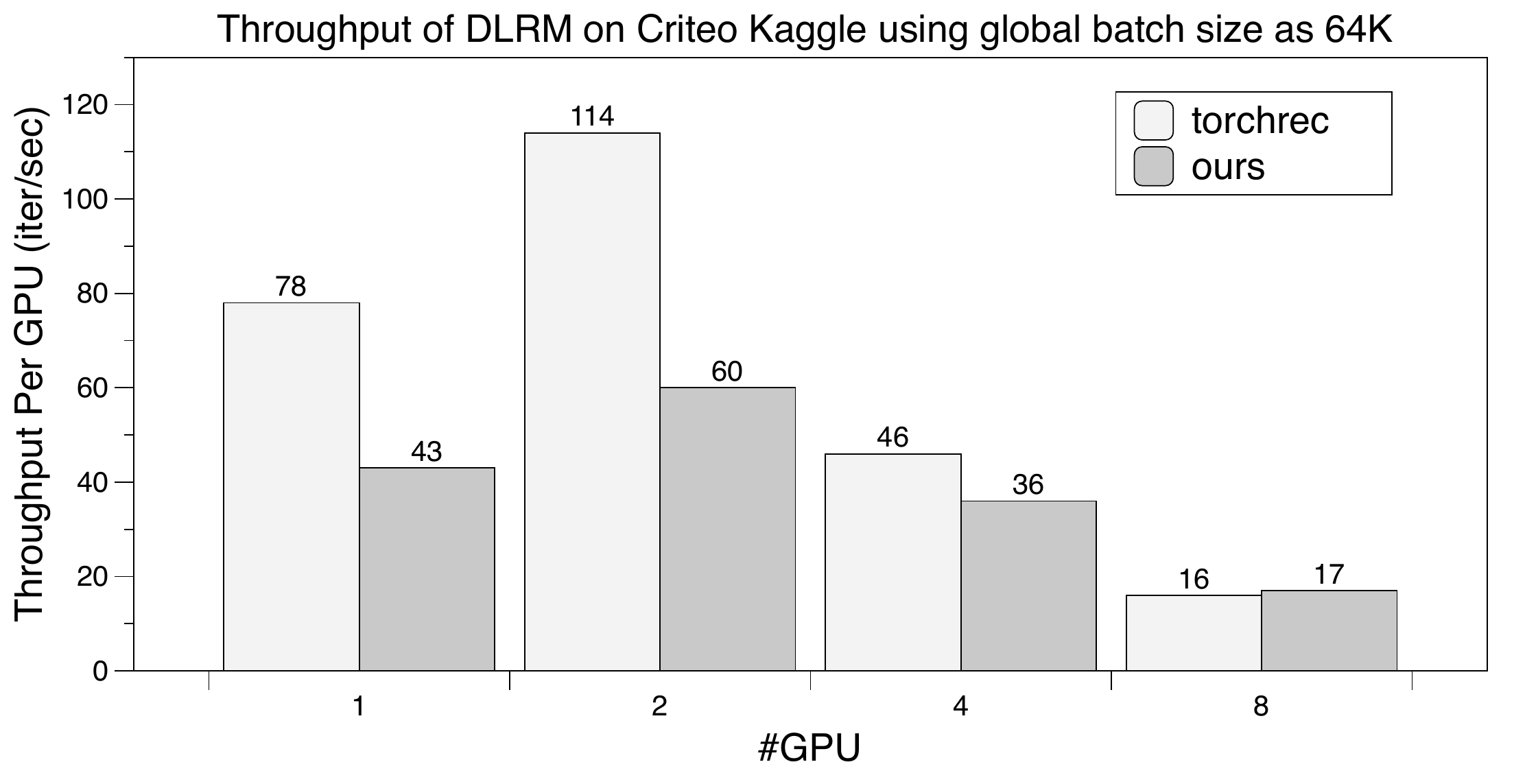}
\caption{Throughput of DLRM on Criteo Kaggle vs. Cache Ratio using batch size as 16K on 1-8 GPUs.}
\label{fig:criteo_mgpu}
\end{figure}

\begin{figure}[ht!]
\centering
\includegraphics[width=0.45\textwidth]{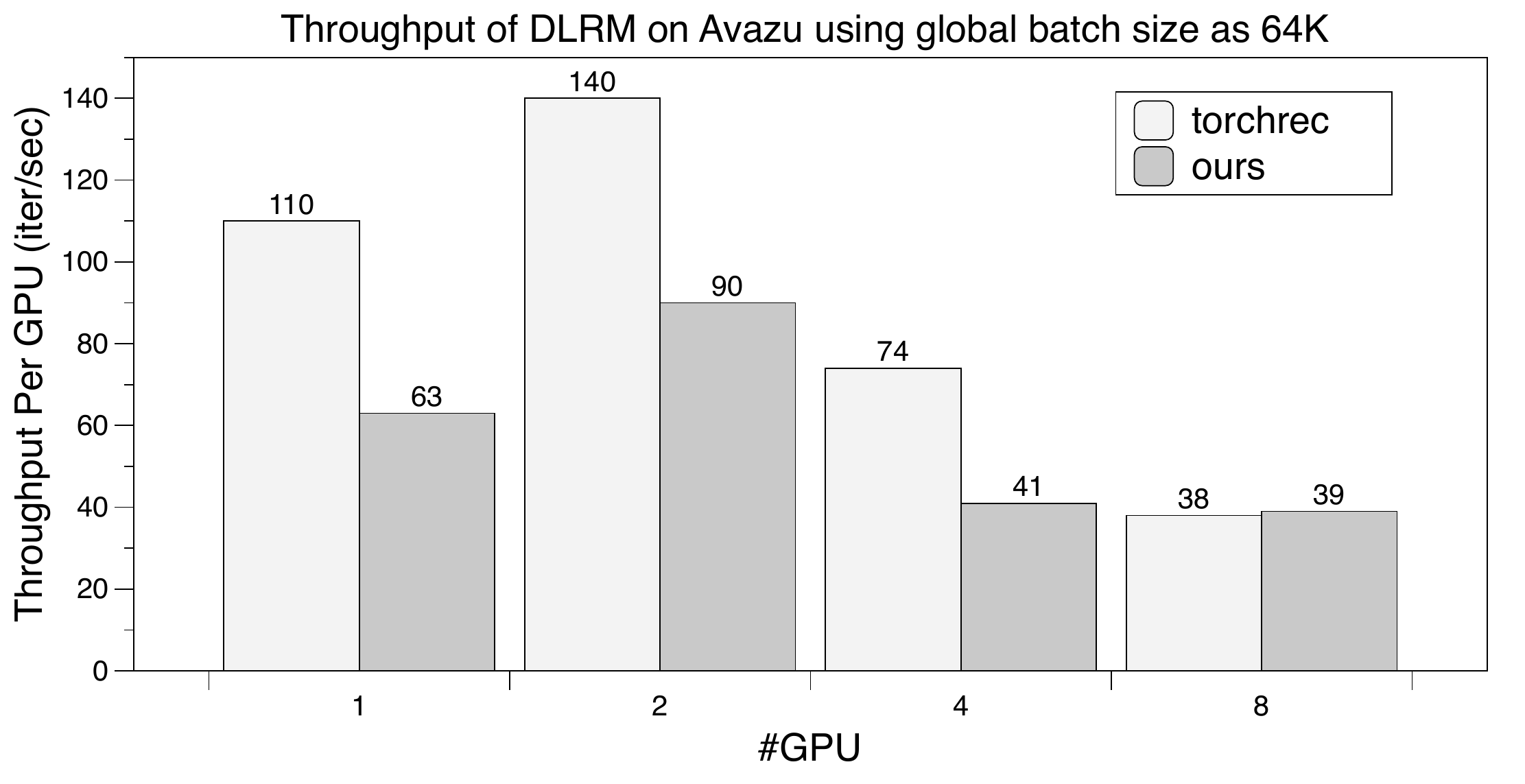}
\caption{Throughput of DLRM on Avazu using global batch size as 64K on 1-8 GPUs.}
\label{fig:avazu_mgpu}
\end{figure}

Figure~\ref{fig:criteo_mgpu} and Figure~\ref{fig:avazu_mgpu} present the throughput when we scale the training from 1 GPU to 8 GPUs.
In our testbed, GPUs are connected by PCI-e and the communication bandwidth between GPUs is limited.
It damages the scalabilities of both systems.
The gap between the throughput of our system and TorchRec is still existing in multiple GPUs. 
However, it gradually shrinks as the number of GPUs increases, because the proportion of inter-GPU communication becomes larger.

\section{Future Work}
In future work, we will evaluate our method on even larger datasets, such as Criteo 1TB dataset\footnote{https://ailab.criteo.com/criteo-1tb-click-logs-dataset/}.
Also, We will adopt an input-id-prefetch method that looks ahead to more input ids to improve the cache eviction efficacy. 
We will also explore the combination of embedding compression algorithm~\cite{ginart2021mixed} with the caching approach, and manage the memory space NVMe in our caching method.

\section{Conclusion}
This paper proposes a software cache design method to solve the problem that the large-scale embedding tables training cannot be executed in GPU.
Based on the target dataset id frequency statistics, our software cache achieves high-speed indexing and data transmission.
Preliminary experiments on our prototype system show that we can obtain a decent end-to-end training speed by keeping only 1.5\% of the embedding parameters on GPU.
This study paves a way for future studies on building extremely large GPU DLRM systems.

\bibliography{my.bib}


\begin{thebibliography}{10}


\ifx \showCODEN    \undefined \def \showCODEN     #1{\unskip}     \fi
\ifx \showDOI      \undefined \def \showDOI       #1{#1}\fi
\ifx \showISBNx    \undefined \def \showISBNx     #1{\unskip}     \fi
\ifx \showISBNxiii \undefined \def \showISBNxiii  #1{\unskip}     \fi
\ifx \showISSN     \undefined \def \showISSN      #1{\unskip}     \fi
\ifx \showLCCN     \undefined \def \showLCCN      #1{\unskip}     \fi
\ifx \shownote     \undefined \def \shownote      #1{#1}          \fi
\ifx \showarticletitle \undefined \def \showarticletitle #1{#1}   \fi
\ifx \showURL      \undefined \def \showURL       {\relax}        \fi
\providecommand\bibfield[2]{#2}
\providecommand\bibinfo[2]{#2}
\providecommand\natexlab[1]{#1}
\providecommand\showeprint[2][]{arXiv:#2}

\bibitem[\protect\citeauthoryear{Adnan, Maboud, Mahajan, and Nair}{Adnan
  et~al\mbox{.}}{2021}]%
        {adnan2021accelerating}
\bibfield{author}{\bibinfo{person}{Muhammad Adnan},
  \bibinfo{person}{Yassaman~Ebrahimzadeh Maboud}, \bibinfo{person}{Divya
  Mahajan}, {and} \bibinfo{person}{Prashant~J Nair}.}
  \bibinfo{year}{2021}\natexlab{}.
\newblock \showarticletitle{Accelerating recommendation system training by
  leveraging popular choices}.
\newblock \bibinfo{journal}{\emph{arXiv preprint arXiv:2103.00686}}
  (\bibinfo{year}{2021}).
\newblock


\bibitem[\protect\citeauthoryear{Adnan, Maboud, Mahajan, and Nair}{Adnan
  et~al\mbox{.}}{2022}]%
        {adnan2022heterogeneous}
\bibfield{author}{\bibinfo{person}{Muhammad Adnan},
  \bibinfo{person}{Yassaman~Ebrahimzadeh Maboud}, \bibinfo{person}{Divya
  Mahajan}, {and} \bibinfo{person}{Prashant~J Nair}.}
  \bibinfo{year}{2022}\natexlab{}.
\newblock \showarticletitle{Heterogeneous Acceleration Pipeline for
  Recommendation System Training}.
\newblock \bibinfo{journal}{\emph{arXiv preprint arXiv:2204.05436}}
  (\bibinfo{year}{2022}).
\newblock


\bibitem[\protect\citeauthoryear{Bian, Liu, Wang, Huang, Li, Wang, Cui, and
  You}{Bian et~al\mbox{.}}{2021}]%
        {bian2021colossal}
\bibfield{author}{\bibinfo{person}{Zhengda Bian}, \bibinfo{person}{Hongxin
  Liu}, \bibinfo{person}{Boxiang Wang}, \bibinfo{person}{Haichen Huang},
  \bibinfo{person}{Yongbin Li}, \bibinfo{person}{Chuanrui Wang},
  \bibinfo{person}{Fan Cui}, {and} \bibinfo{person}{Yang You}.}
  \bibinfo{year}{2021}\natexlab{}.
\newblock \showarticletitle{Colossal-AI: A Unified Deep Learning System For
  Large-Scale Parallel Training}.
\newblock \bibinfo{journal}{\emph{arXiv preprint arXiv:2110.14883}}
  (\bibinfo{year}{2021}).
\newblock


\bibitem[\protect\citeauthoryear{Ginart, Naumov, Mudigere, Yang, and
  Zou}{Ginart et~al\mbox{.}}{2021}]%
        {ginart2021mixed}
\bibfield{author}{\bibinfo{person}{Antonio~A Ginart}, \bibinfo{person}{Maxim
  Naumov}, \bibinfo{person}{Dheevatsa Mudigere}, \bibinfo{person}{Jiyan Yang},
  {and} \bibinfo{person}{James Zou}.} \bibinfo{year}{2021}\natexlab{}.
\newblock \showarticletitle{Mixed dimension embeddings with application to
  memory-efficient recommendation systems}. In \bibinfo{booktitle}{\emph{2021
  IEEE International Symposium on Information Theory (ISIT)}}. IEEE,
  \bibinfo{pages}{2786--2791}.
\newblock


\bibitem[\protect\citeauthoryear{Gomez-Uribe and Hunt}{Gomez-Uribe and
  Hunt}{2015}]%
        {gomez2015netflix}
\bibfield{author}{\bibinfo{person}{Carlos~A Gomez-Uribe} {and}
  \bibinfo{person}{Neil Hunt}.} \bibinfo{year}{2015}\natexlab{}.
\newblock \showarticletitle{The netflix recommender system: Algorithms,
  business value, and innovation}.
\newblock \bibinfo{journal}{\emph{ACM Transactions on Management Information
  Systems (TMIS)}} \bibinfo{volume}{6}, \bibinfo{number}{4}
  (\bibinfo{year}{2015}), \bibinfo{pages}{1--19}.
\newblock


\bibitem[\protect\citeauthoryear{Lian, Yuan, Zhu, Wang, He, Wu, Sun, Lyu, Liu,
  Dong, et~al\mbox{.}}{Lian et~al\mbox{.}}{2021}]%
        {lian2021persia}
\bibfield{author}{\bibinfo{person}{Xiangru Lian}, \bibinfo{person}{Binhang
  Yuan}, \bibinfo{person}{Xuefeng Zhu}, \bibinfo{person}{Yulong Wang},
  \bibinfo{person}{Yongjun He}, \bibinfo{person}{Honghuan Wu},
  \bibinfo{person}{Lei Sun}, \bibinfo{person}{Haodong Lyu},
  \bibinfo{person}{Chengjun Liu}, \bibinfo{person}{Xing Dong}, {et~al\mbox{.}}}
  \bibinfo{year}{2021}\natexlab{}.
\newblock \showarticletitle{Persia: a hybrid system scaling deep learning based
  recommenders up to 100 trillion parameters}.
\newblock \bibinfo{journal}{\emph{arXiv preprint arXiv:2111.05897}}
  (\bibinfo{year}{2021}).
\newblock


\bibitem[\protect\citeauthoryear{Miao, Zhang, Shi, Nie, Yang, Tao, and
  Cui}{Miao et~al\mbox{.}}{2021}]%
        {miao2021het}
\bibfield{author}{\bibinfo{person}{Xupeng Miao}, \bibinfo{person}{Hailin
  Zhang}, \bibinfo{person}{Yining Shi}, \bibinfo{person}{Xiaonan Nie},
  \bibinfo{person}{Zhi Yang}, \bibinfo{person}{Yangyu Tao}, {and}
  \bibinfo{person}{Bin Cui}.} \bibinfo{year}{2021}\natexlab{}.
\newblock \showarticletitle{Het: Scaling out huge embedding model training via
  cache-enabled distributed framework}.
\newblock \bibinfo{journal}{\emph{arXiv preprint arXiv:2112.07221}}
  (\bibinfo{year}{2021}).
\newblock


\bibitem[\protect\citeauthoryear{Mudigere, Hao, Huang, Jia, Tulloch, Sridharan,
  Liu, Ozdal, Nie, Park, et~al\mbox{.}}{Mudigere et~al\mbox{.}}{2022}]%
        {mudigere2022software}
\bibfield{author}{\bibinfo{person}{Dheevatsa Mudigere}, \bibinfo{person}{Yuchen
  Hao}, \bibinfo{person}{Jianyu Huang}, \bibinfo{person}{Zhihao Jia},
  \bibinfo{person}{Andrew Tulloch}, \bibinfo{person}{Srinivas Sridharan},
  \bibinfo{person}{Xing Liu}, \bibinfo{person}{Mustafa Ozdal},
  \bibinfo{person}{Jade Nie}, \bibinfo{person}{Jongsoo Park}, {et~al\mbox{.}}}
  \bibinfo{year}{2022}\natexlab{}.
\newblock \showarticletitle{Software-hardware co-design for fast and scalable
  training of deep learning recommendation models}. In
  \bibinfo{booktitle}{\emph{Proceedings of the 49th Annual International
  Symposium on Computer Architecture}}. \bibinfo{pages}{993--1011}.
\newblock


\bibitem[\protect\citeauthoryear{Naumov, Mudigere, Shi, Huang, Sundaraman,
  Park, Wang, Gupta, Wu, Azzolini, et~al\mbox{.}}{Naumov et~al\mbox{.}}{2019}]%
        {naumov2019deep}
\bibfield{author}{\bibinfo{person}{Maxim Naumov}, \bibinfo{person}{Dheevatsa
  Mudigere}, \bibinfo{person}{Hao-Jun~Michael Shi}, \bibinfo{person}{Jianyu
  Huang}, \bibinfo{person}{Narayanan Sundaraman}, \bibinfo{person}{Jongsoo
  Park}, \bibinfo{person}{Xiaodong Wang}, \bibinfo{person}{Udit Gupta},
  \bibinfo{person}{Carole-Jean Wu}, \bibinfo{person}{Alisson~G Azzolini},
  {et~al\mbox{.}}} \bibinfo{year}{2019}\natexlab{}.
\newblock \showarticletitle{Deep learning recommendation model for
  personalization and recommendation systems}.
\newblock \bibinfo{journal}{\emph{arXiv preprint arXiv:1906.00091}}
  (\bibinfo{year}{2019}).
\newblock


\bibitem[\protect\citeauthoryear{Smith and Linden}{Smith and Linden}{2017}]%
        {smith2017two}
\bibfield{author}{\bibinfo{person}{Brent Smith} {and} \bibinfo{person}{Greg
  Linden}.} \bibinfo{year}{2017}\natexlab{}.
\newblock \showarticletitle{Two decades of recommender systems at Amazon. com}.
\newblock \bibinfo{journal}{\emph{Ieee internet computing}}
  \bibinfo{volume}{21}, \bibinfo{number}{3} (\bibinfo{year}{2017}),
  \bibinfo{pages}{12--18}.
\newblock


\end{thebibliography}
\bibliographystyle{ACM-Reference-Format}

\end{document}